\documentclass[9pt,twocolumn,twoside]{osajnl}

\journal{ol} 

\setboolean{shortarticle}{true}

\title{Non-imaging single-pixel sensing with optimized binary modulation}

\author[1,2]{Hao Fu}
\author[1,2,*]{Liheng Bian}
\author[1,2]{Jun Zhang}

\affil[1]{School of Information and Electronics, Beijing Institute of Technology, Beijing, 100081}
\affil[2]{Advanced Research Institute of Multidisciplinary Science, Beijing Institute of Technology, Beijing, 100081}

\affil[*]{bian@bit.edu.cn}

\begin{abstract}
The conventional high-level sensing techniques require high-fidelity images as input to extract target features, which are produced by either complex imaging hardware or high-complexity reconstruction algorithms. In this letter, we propose single-pixel sensing (SPS) that performs high-level sensing directly from coupled measurements of a single-pixel detector, without the conventional image acquisition and reconstruction process. The technique consists of three steps including binary light modulation that can be physically implemented at $\sim$22kHz, single-pixel coupled detection owning wide working spectrum and high signal-to-noise ratio, and end-to-end deep-learning based sensing that reduces both hardware and software complexity. Besides, the binary modulation is trained and optimized together with the sensing network, which ensures least required measurements and optimal sensing accuracy. The effectiveness of SPS is demonstrated on the classification task of handwritten MNIST dataset, and 96.68\% classification accuracy at $\sim$1kHz is achieved. The reported single-pixel sensing technique is a novel framework for highly efficient machine intelligence.
\end{abstract}

\setboolean{displaycopyright}{true}

\begin{document}

\maketitle

\section{Introduction}

As the development of computer vision and computer graphics, there rises a number of high-level sensing tasks that teach machines to perceive the physical world, such as image classification, semantic segmentation and object detection \cite{forsyth2002computer}. These applications require high-fidelity images as input to extract precise target features, which largely rely on complex imaging hardware such as high-sensitivity sensors and achromatic lens, and high-complexity reconstruction software such as denoising and deblurring algorithms \cite{russ2016image}. These requirements make current sensing systems with high cost, much running time, low sensing rate and heavy communication load. The drawbacks impede the promotion and popularization of practical high-level sensing applications.


\begin{figure}[t]
\label{fig1}
\centering
\centerline{\includegraphics[width=\linewidth]{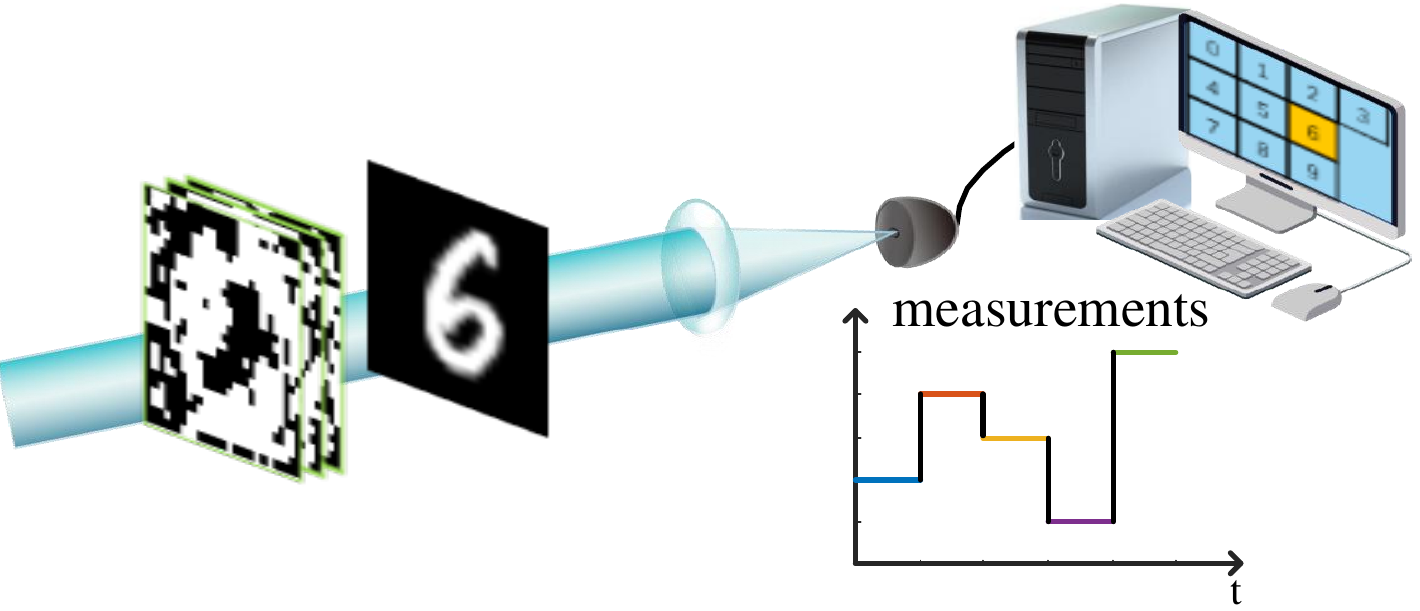}}
\caption{The system schematic of the reported single-pixel sensing (SPS) technique.}
\label{fig1}
\end{figure}

In most applications, however, the target locates only at partial regions of interests instead of the entire field of view. In other words, there exist a certain number of pixels that contain little useful information for sensing in natural images and videos. These informationless pixels waste both hardware and software costs. In this sense, the conventional imaging-sensing framework is not an optimal choice for machine intelligence to a certain extent. Intuitively, bypassing the complex imaging process and directly obtaining target features instead of images/videos is a feasible solution to improve the efficiency of both information acquisition and sensing.

To this end, we propose a novel technique termed single-pixel sensing (SPS) in this letter. SPS performs high-level sensing directly from coupled measurements of a single-pixel detector instead of an array sensor, without the conventional image acquisition and reconstruction process. SPS shares its roots with the single-pixel imaging (SPI) technique \cite{duarte2008single, edgar2018principles, bian2018experimental}. In the conventional SPI system, the light is first modulated by a spatial light modulator, and then converged and captured by a single-pixel detector. Optimization algorithms such as compressive sensing methods are required to reconstruct the target image, from the coupled measurement sequence and corresponding modulation patterns. Differently in the proposed SPS technique, however, the data is directly input into an end-to-end neural network for high-level sensing. Therefore, the conventional image reconstruction process is removed, and the computational complexity is largely decreased. 

The compressed learning (CL) technique is also related to the proposed SPS method. Davenport et al. first proposed the concept of compressed classification in 2007 that utilizes random binary patterns to modulate light, and derived a maximum likelihood optimization algorithm to classify the target scene from compressed measurements \cite{davenport2007smashed}. The strategy was then extended to compressed learning that enables to deal with various high-level visual inference tasks such as face recognition and action recognition, using either the conventional statistical optimization methods \cite{lohit2015reconstruction, kulkarni2015reconstruction, vargas2018object} or the popular deep learning methods \cite{lohit2016direct, xu2019compressed, wang2019privacy}. However, the modulation patterns utilized in the above methods are randomly generated. Though random modulation is able to maintain irrelevance among different measurements, it may be not an optimal choice owning the highest efficiency for a specific classification task or a specific class of target scenes \cite{adler2016compressed}. To optimize light modulation, Adler et al. \cite{adler2016compressed} took modulation patterns as a part of a neural network, which were jointly optimized with the subsequent inference operator. However, the pattern training process is implemented with no specific regularization. As a result, the modulation patterns are in gray-scale and contain negative values, which are hard or need much time to physically implement using a spatial light modulator. Consequently, both the hardware complexity and acquisition time are largely increased.

To increase modulation efficiency, we build an end-to-end deep learning framework to optimize the binary modulation patterns. The framework consists of both an encoder that modulates and couples target light field into one-dimensional measurements, and a decoder that infers semantic information from the single-pixel coupled measurements. The weights of the encoder are regularized to be binary in the training process, which is beneficial for fast physical implementation ($\sim$22kHz). The encoder and decoder are trained together as a whole neural network, which ensures both least required measurements and best sensing performance. Once trained, the encoder weights represent the binary modulation patterns to generate coupled measurements, and correspondingly the decoder is able to output final sensing results directly from the coupled measurements in an end-to-end manner. This largely reduces computational complexity and storage memory cost.

Since the SPS system employs single-pixel detection, it inherits the advantages of SPI systems in two aspects. First, the working spectrum is widely extended benefiting from the utilized single-pixel detector. This paves the way of non-imaging sensing beyond the visible range where array sensors are expensive or not available. Second, the signal-to-noise ratio is largely increased by orders of magnitude, because all the light from target scene is converged to the single detection unit. This is beneficial for low-light applications. 

\section{Method}

The reported SPS system is sketched in Fig. \ref{fig1}. The light is modulated by a spatial light modulator following the pre-trained binary patterns. Here we employ a digital micromirror device (DMD, ViALUX V-7001) that works at $\sim$22kHz in the binary modulation mode, which is two orders-of-magnitude faster than the conventional gray-scale modulation. The modulated light is conjugated to the target scene, and the transmitted/reflected light is then converged by a lens to a single-pixel detector. We employ an Si amplified photodetector (Thorlabs PDA100A2, 320-1100nm) together with a 14-bit acquisition board (ART PCI8514) in the proof-of-concept setup. The detector acquires the total light intensity. With a series of patterns varying by the DMD, a sequence of coupled measurements is obtained. Different from the conventional SPI system that utilizes both the acquired data and modulation patterns to first reconstruct the target image and then extract features for subsequent sensing, the SPS system directly inputs only the acquired data into the pre-trained end-to-end decoder network and outputs the final sensing results.


\begin{figure}[!t]
\label{fig2}
\centering
\centerline{\includegraphics[width=\linewidth]{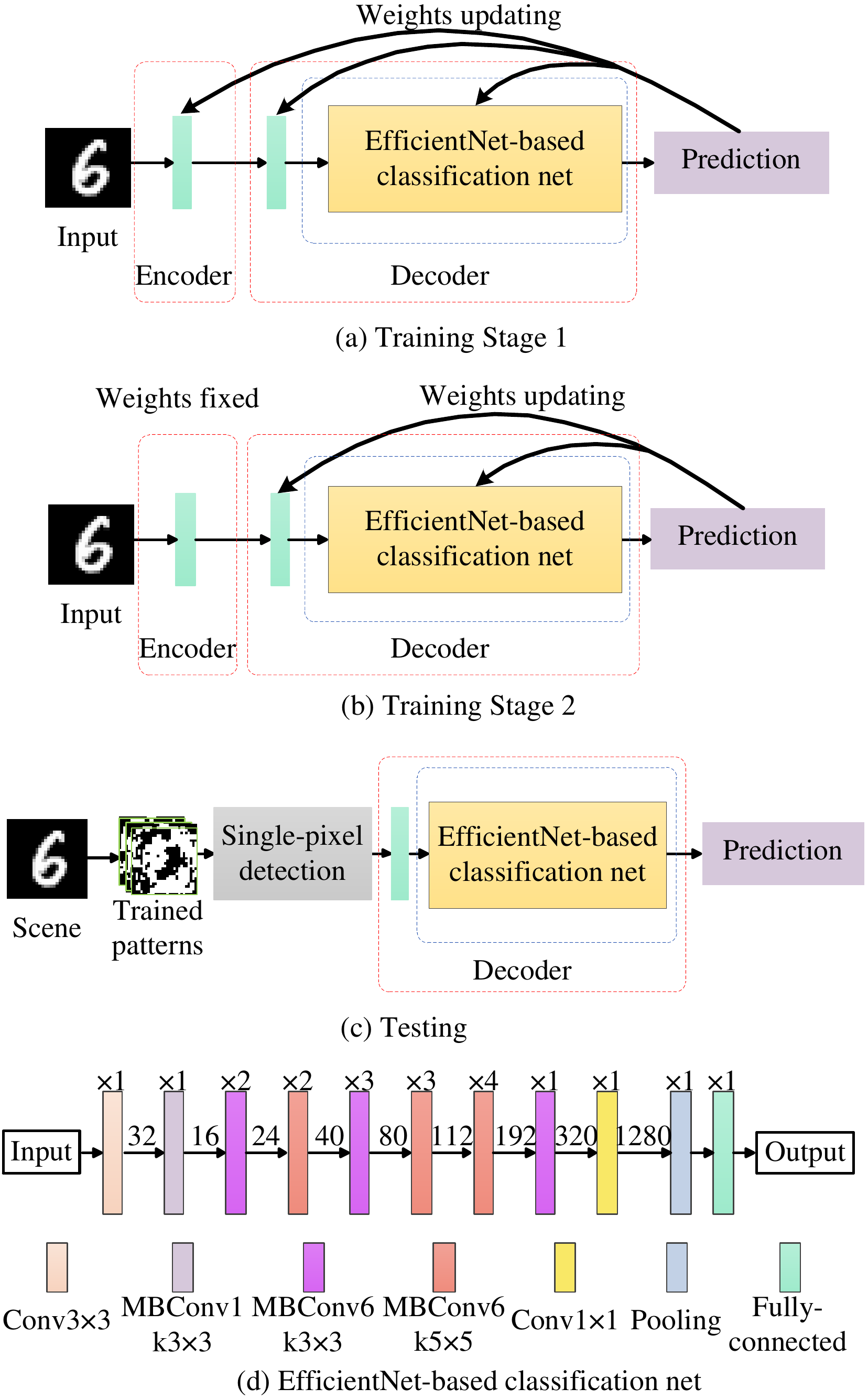}}
\caption{Overview of the reported SPS network. (a) and (b) show the two training stages of the network, (c) is the testing operation, and (d) presents the structure of the designed EfficientNet-based decoder network. MBConv denotes the mobile inverted bottleneck convolution layer, k3×3/k5×5 represents kernel size, ×1/2/3/4 denotes the number of repeated layers within the block. The digits above the arrows indicate the number of channels.}
\label{fig2}
\end{figure}

The designed end-to-end sensing network is shown in Fig. \ref{fig2}. It consists of two parts including the target encoder for binary modulation with high speed, and the sensing decoder to produce sensing results from the coupled measurements with high accuracy. The encoder contains one fully-connected layer without bias representing the modulation patterns. Assuming that the pixel number of target scene is M × N, the number of input features of the fully-connected layer is M × N, and the number of output features is M × N × R (R is sampling ratio defined as the ratio between measurement number and pixel number). The sensing decoder comprises two parts. The first part has one fully-connected layer for feature inference, with the number of input features being M × N × R and the number of output features being M × N. 
The second part outputs final sensing results. Different sensing tasks require different sensing networks to produce best performance. Here in this letter, we take the image classification as an example, and the sensing network is designed based on the state-of-the-art image classification network structure termed EfficientNet \cite{tan2019efficientnet}. EfficientNet is a convolutional neural network, whose structure is optimized by searching and balancing network depth, width and resolution. The designed EfficientNet-based classification network is shown in Fig. \ref{fig2}, which contains 11 convolution stages in total that are comprised of different types of the mobile inverted bottleneck MBConv \cite{sandler2018mobilenetv2, tan2019mnasnet}.

The network training process consists of two stages, as shown in Fig. \ref{fig2}(a) and (b). In stage 1, the weights of both the encoder and decoder are optimized simultaneously. In practice, the weights of the encoder is gray-scale and contains negative values if there is no ragularization. To realize binary modulation, an extra binarizing operation on the encoder weights is required for each epoch of forward propagation. The metric of optimized binary weights is to make them as close as possible to the gray-scale ones to ensure minimum loss. Mathmatically, assuming that there exists a scaling factor $\alpha$ so that 
\begin{equation}
\label{eq1}
\mathbf{W} \approx \alpha \mathbf{W}_{b},
\end{equation}
where $\mathbf{W}$ is the gray-scale weight matrix, and $\alpha\mathbf{W}_{b} \in\{\alpha,0\}$ denotes its binarized version calculated as
\begin{equation}
\label{eq5}
\alpha\mathbf{W}_{b}=\alpha\operatorname{sign}(\mathbf{W})=\left\{\begin{array}{ll}{\alpha} & {\mathbf{W} > 0} \\ {0} & {\text { otherwise }}.
\end{array}\right.
\end{equation} 
Based on the above notations, obtaining the optimal binary weights equals to solving the following optimization problem as \cite{rastegari2016xnor}
\begin{equation}
\label{eq2}
\alpha^{*}=\underset{\alpha}{\operatorname{argmin}}\left\|\mathbf{W}-\alpha \mathbf{W}_{{b}}\right\|^{2},
\end{equation}
which can be rewritten as
\begin{equation}
\label{eq3}
\underset{\alpha}{\operatorname{argmin}}~F(\alpha)=\alpha^{2} \mathbf{W}_{b}^{\top} \mathbf{W}_{b}-2 \alpha \mathbf{W}^{\top} \mathbf{W}_{b}+\mathbf{W}^{\top} \mathbf{W}.
\end{equation}
Taking the derivative of $F(\alpha)$ with respect to $\alpha$ and setting it to zero, we have
\begin{equation}
\label{eq6}
\alpha^{*}=\frac{\mathbf{W}^{\top} \mathbf{W}_{b}}{\mathbf{W}_{b}^{\top} \mathbf{W}_{b}}.
\end{equation}

\begin{figure}[!t]
\centering
\centerline{\includegraphics[width=\linewidth]{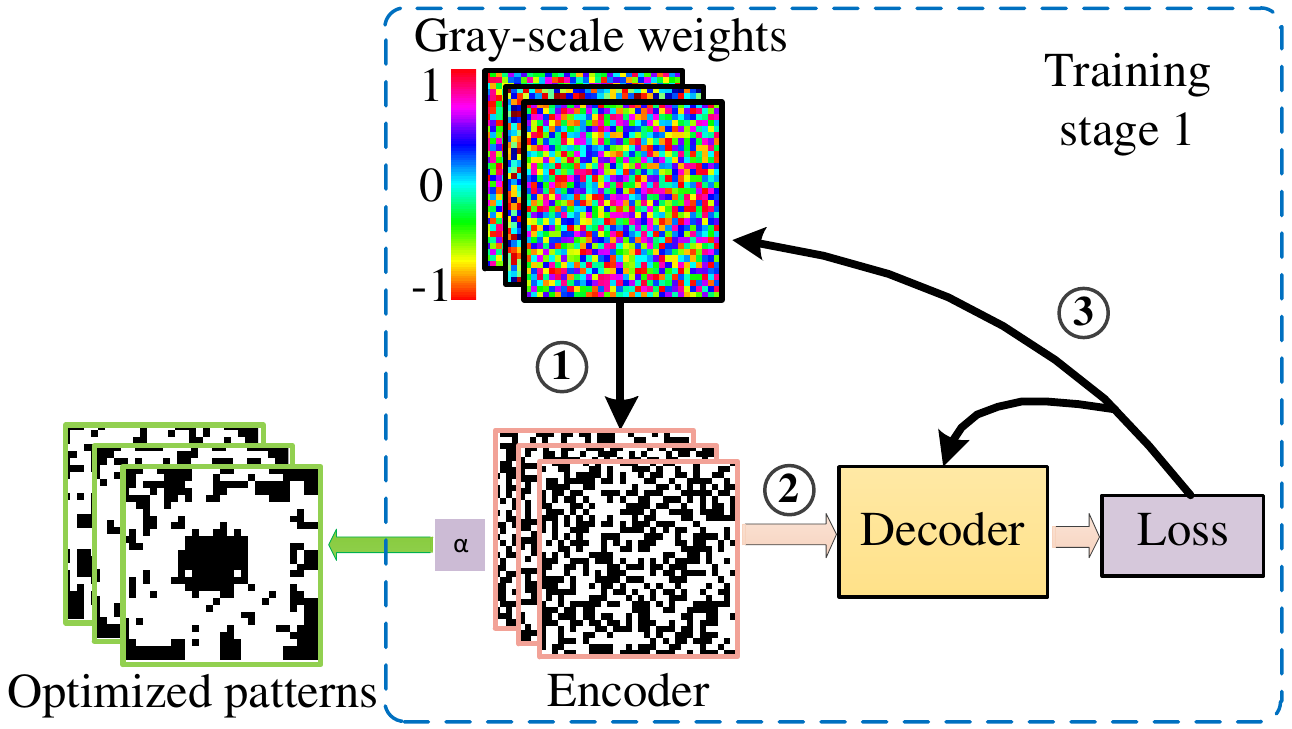}}
\caption{Overview of the first training stage (3 steps). Step 1: binarize the gray-scale weights and calculate the scaling factor $\alpha$. Step 2: perform the forward propagation. Step 3: perform the back propagation and update the gray-scale weights. Finally, the optimized binary patterns are obtained after the training stage 1.}
\label{fig3}
\end{figure}

For each forward propagation, as shown in Fig. \ref{fig3}, the scaling factor $\alpha$ and binarized weights $\alpha\mathbf{W}_{b}$ are first calculated following Eq. \ref{eq6} and Eq. \ref{eq5}. Then, the gray-scale weights are replaced with those binarized ones, which are utlized to calculate the network loss and corresponding gradients. In each backward propagation, the gray-scale weights are updated using the gradients. In such an iterative training strategy, the optimal binary encoder weights can be obtained.

To further improve sensing accuracy, we design the additional training stage 2 as shown in Fig. \ref{fig2} (b), in which the encoder weights are fixed and only the decoder weights are updated. The reason of incorporating an additional training stage is that at the end of training stage 1, though the gradient flow goes back through the whole network, the little variation of the gray-scale encoder weights ($\mathbf{W}$) is not enough to inverse the sign of its weights ($\mathbf{W}_{b}$). In such case, the binarized encoder continues to bring loss to the framework and impedes its optimization. To avoid the updating stagnation, the encoder weights are fixed as its binary version, and only the decoder weights are updated. In this way, the gradient flow only goes through the decoder to better match the fixed encoder and improve sensing performance.
 
In the testing part as shown in Fig. \ref{fig2} (c), the optimized binary weights of the encoder are resized as the light modulation patterns. Correspondingly, the single-pixel detector acquires a sequence of coupled measurements, which are fed as the input of the sensing decoder. The decoder outputs the sensing results of the target scene.

\section{Results}

In order to evaluate the performance of the reported SPS technique, we implement the method on the classification task of the hand-written MNIST dataset \cite{lecun1998gradient}. The MNIST dataset comprises 60000 training images and 10000 testing images, which spread over 10 classes from digit `0' to `9' in total. Each image is in gray scale and contains 28 × 28 pixels. In the training stage 1, we use the Adam solver for gradient optimization, and set the weight decay being 1e-4. The learning rate is initialized as 1e-3, and is decreased by a factor of 10 for every 30 epochs. A total number of 80 epochs are trained. The loss function used here is cross-entropy. In the training stage 2, we fix the encoder weights, and only update the decoder weights with the same training parameter settings as stage 1. 

\begin{table}[b]
\label{tab1}
\caption{Classification accuracy on the MNIST dataset under different sampling ratios.}
\begin{tabular}{@{}ccc@{}}
\toprule
Sampling ratio & Measurement number & Accuracy(\%) \\ \midrule
0.01           & 8                      & 91.73        \\
0.03           & 22                     & 96.68        \\
0.05           & 39                     & 97.14        \\
0.1            & 78                     & 97.36        \\ \bottomrule
\end{tabular}
\end{table}

The classification performance on the MNIST dataset is shown in Table 1. By defining the sampling ratio as the ratio between measurement number and pixel number, we explore the influence of different sampling ratios on final classification accuracy.  In experiments, we set the sampling ratio ranging from 0.01 to 0.1, and corresponding measurement number ranges from 8 to 78. The results show that the classification accuracy improves as the sampling ratio increases. Specifically, the classification accuracy increases from 91.73\% to 96.68\% as sampling ratio increases from 0.01 to 0.03. As sampling ratio further increases above 0.03, the classification accuracy improves little. To conclude, benefiting from the binary modulation and joint-training framework, the classification can be implemented at the rate of as fast as $\sim$1kHz, with $\sim$97\% classification accuracy.

\section{Conclusion and discussion}

In this letter, we report a non-imaging single-pixel sensing (SPS) technique that performs high-level sensing without images. The technique consists of three steps including binary light modulation, single-pixel coupled detection, and end-to-end sensing. The advantages of the SPS technique lie in the following three aspects. First, it removes the conventional image acquisition and reconstruction process, and utilizes an end-to-end neural network to directly perceive the target from coupled measurements, which reduces both hardware and software complexity. Second, the binary modulation is trained and optimized together with the sensing network, which ensures least required measurements, fast physical implementation and high sensing rate. Third, the single-pixel detection brings wide working spectrum and high signal-to-noise ratio, which benefit for non-visible-range and low-light applications. The SPS technique has been successfully demonstrated on the classification task of handwritten MNIST dataset, and experiment results show that it enables 96.68\% classification accuracy that can be performed at $\sim$1kHz.

The SPS technique is beneficial for communication and encryption. The reasons lie in the following two aspects. First, the required data contains only the coupled single-pixel measurements, which is around two orders-of-magnitude less than the target image. Such a small data size largely relieves communication load. Second, the decoder matches only to the optimized encoder because both of them are trained as a whole neural network. In this sense, there is little chance to reproduce the decoder without modulation information, and thus the mechanism is beneficial for encryption.

The SPS technique can be further extended. First, the network structures of both the encoder and decoder can be further investigated and improved. This may further reduce modulation patterns and improve sensing performance. Second, the pixel resolution of light modulation can be dynamically adapted for different target scenes and different sensing tasks. For those applications that require fewer features, neighboring pixels can be binned together to reduce modulation patterns while ensuring sensing  accuracy.

\section{Funding Information}

Fundamental Research Funds for the Central Universities (Grant No. 3052019024); National Natural Science Foundation of China (Grant No. 61971045, 61827901).

\section{Disclosures}

The authors declare no conflicts of interest.

\bibliography{sample}

\bibliographyfullrefs{sample}


\ifthenelse{\equal{\journalref}{aop}}{%
\section*{Author Biographies}
\begingroup
\setlength\intextsep{0pt}
\begin{minipage}[t][6.3cm][t]{1.0\textwidth} 
  \begin{wrapfigure}{L}{0.25\textwidth}
    \includegraphics[width=0.25\textwidth]{john_smith.eps}
  \end{wrapfigure}
  \noindent
  {\bfseries John Smith} received his BSc (Mathematics) in 2000 from The University of Maryland. His research interests include lasers and optics.
\end{minipage}
\begin{minipage}{1.0\textwidth}
  \begin{wrapfigure}{L}{0.25\textwidth}
    \includegraphics[width=0.25\textwidth]{alice_smith.eps}
  \end{wrapfigure}
  \noindent
  {\bfseries Alice Smith} also received her BSc (Mathematics) in 2000 from The University of Maryland. Her research interests also include lasers and optics.
\end{minipage}
\endgroup
}{}

\end{document}